\documentclass[aps,prl,secnumarabic,nobibnotes,twocolumn,superscriptaddress,longbibliography]{revtex4-2}
\usepackage{amsfonts}
\usepackage{mathrsfs}
\usepackage{amsmath}
\usepackage{color}
\usepackage{natbib}
\usepackage{graphicx}
\usepackage{bm}
\usepackage{amssymb}
\usepackage{xspace}
\usepackage{epstopdf}
\usepackage{dcolumn}
\usepackage{multirow}
\usepackage[colorlinks=true, letterpaper=true, pdfstartview=FitV, linkcolor=blue, citecolor=blue, urlcolor=blue]{hyperref}
\usepackage{wrapfig}

\makeatletter

\newcommand{\Rmnum}[1]{\expandafter\@slowromancap\romannumeral #1@}
\makeatother

\begin{document}
\title{Electric-Field Control of Terahertz Response via Spin--Corner--Layer Coupling in Altermagnetic Bilayers}

\author{Jianhua Wang}
\address{Institute for Superconducting and Electronic Materials, Faculty of Engineering and Information Sciences, University of Wollongong, Wollongong 2500, Australia.}
\address{School of Material Science and Engineering, Tiangong University, Tianjin 300387, China}

\author{Yilin Han}
\address{Key Lab of Advanced Optoelectronic Quantum Architecture and Measurement (MOE), Beijing Key Lab of Nanophotonics $\&$ Ultrafine Optoelectronic Systems, and School of Physics, Beijing Institute of Technology, Beijing 100081, China.}

\author{Shifeng Qian}\thanks{Corresponding authors}\email{qiansf@ahnu.edu.cn}
\address{Anhui Province Key Laboratory for Control and Applications of Optoelectronic Information Materials, Department of Physics, Anhui Normal University, Wuhu, Anhui 241000, China.}

\author{Zhenxiang Cheng}\thanks{Corresponding authors}\email{cheng@uow.edu.au}
\address{Institute for Superconducting and Electronic Materials, Faculty of Engineering and Information Sciences, University of Wollongong, Wollongong 2500, Australia.}
\author{Wenhong Wang}
\address{School of Material Science and Engineering, Tiangong University, Tianjin 300387, China}

\author{Zhi-Ming Yu}
\address{Key Lab of Advanced Optoelectronic Quantum Architecture and Measurement (MOE), Beijing Key Lab of Nanophotonics $\&$ Ultrafine Optoelectronic Systems, and School of Physics, Beijing Institute of Technology, Beijing 100081, China.}

\author{Xiaotian Wang}\thanks{Corresponding authors}\email{xiaotianw@uow.edu.au}
\address{Institute for Superconducting and Electronic Materials, Faculty of Engineering and Information Sciences, University of Wollongong, Wollongong 2500, Australia.}

\begin{abstract}
Electric field control of electron charge and spin degrees of freedom is fundamental to modern semiconductor and spintronic devices. Yet controlling electromagnetic waves with an electric field, particularly in the terahertz (THz) band, remains a challenge. Here, we propose a spin-corner-layer coupling (SCLC) mechanism in second-order topological altermagnetic bilayers. By using an electric field to influence electrons between different layers, the SCLC mechanism enables simultaneous control over corner and spin degrees of freedom, thereby allowing electric-field tuning of the absorption,  emission intensity, and even polarization of THz waves. Taking bilayer NiZrI$_6$ nanodisks as a prototype, we demonstrate that an ultralow electrostatic field can switch both the spin and the layer polarizations of corner states. This dual switching modulates transition dipole moments and oscillator strengths between different corner states, thereby enabling the manipulation of THz waves. This study establishes a mechanism for the electric-field control of spin and THz waves through SCLC, yielding important implications for the advancement of THz spintronics.

\end{abstract}

\maketitle

\textcolor{blue}{\textit{Introduction}}---
Terahertz (THz) waves~\cite{PhysRevLett.71.2725,PhysRevE.49.671,PhysRevLett.96.075005,PhysRevLett.109.243002,0} have attracted significant attention due to their broad applications in high-speed wireless communication~\cite{ma2018security}, non-intrusive standoff sensing~\cite{ding2010high,liu2010broadband}, and industrial quality control~\cite{rutz2006terahertz,duling2009revealing}. Therefore, achieving effective control over THz waves holds great scientific significance and application value. Electric fields have long been established as the most direct and mature external tool in semiconductor electronics and spintronics, where they have been successfully used to manipulate charge and spin degrees of freedom (d.o.f.), forming the foundation of the entire information industry. Thus, a natural strategy is to channel electric-field control of spin and related d.o.f. into controlling THz electromagnetic waves. Consistent with this strategy, the emerging field of THz spintronics~\cite{9,10,1,11,2,3} exploits the electron's intrinsic spin to overcome the limitations of conventional electronics and optics and is now a rapidly expanding research frontier.

\begin{figure}
\includegraphics[width=8.8cm]{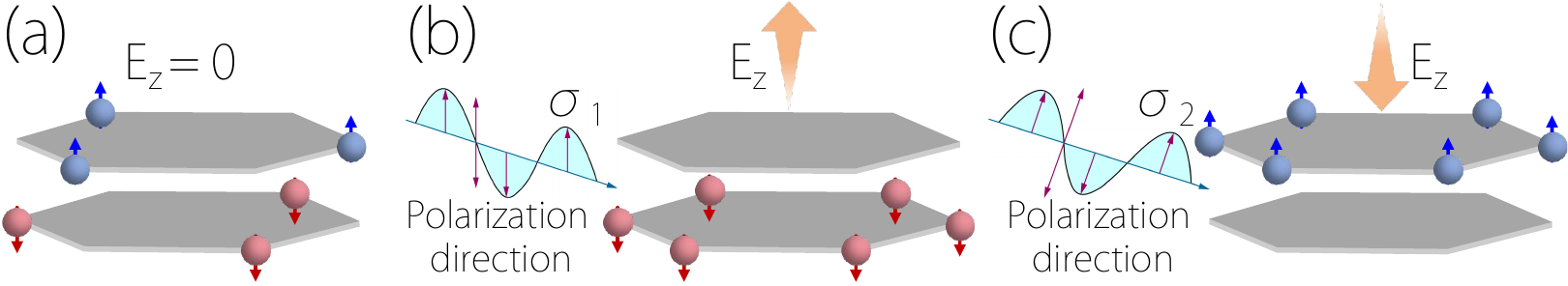}
\caption{Schematic illustration of electric-field control of terahertz response via SCLC in altermagnetic bilayer SOTI nanodisk. (a) In the absence of an out-of-plane electric field ($E_z$), the system exhibits intrinsic SCLC. (b,c) By reversing the direction of the $E_z$, the system exhibits dual switching of layer polarization ($P_l$) and spin polarization ($P_s$) at the corners, enabling electric-field control of the absorption, emission intensity, and polarization of THz waves. $\sigma_1$ and $\sigma_2$ represent two different polarizations of THz waves.
\label{fig1}}
\end{figure}

Recently, 2D second-order topological insulators (SOTIs)~\cite{13,18,19,20,16,17,12,14,15,21,24,22,23} have attracted attention for their symmetry-protected corner states, which localize at the intersection of two crystal edges. The nanodisks constructed from 2D SOTI act as quantum dots (QDs), possessing a new kind of d.o.f.---corner d.o.f.~\cite{25,26,27,28}---which arises from the corner states and can be controlled by external fields~\cite{25,29}. On the other hand, altermagnetic materials~\cite{34,38,30,35,36,37,guo2025spin,39,40,41,31,32,33,43,44,45,46,47}, distinct from conventional ferromagnets and antiferromagnets, have zero net magnetization but feature spin-split electronic structures. Recent breakthroughs have further shown that electrical manipulation of spin d.o.f. in altermagnetic bilayers is feasible~\cite{48,50,49,51}, providing a platform for achieving electric-field control of spin. Therefore, integrating the corner and layer d.o.f. of SOTI bilayer nanodisks with the intrinsic spin d.o.f. of altermagnets leads to altermagnetic bilayer SOTI nanodisks, a promising platform for electric-field manipulation of spin and corners. Furthermore, corner–state transitions may give rise to THz excitations with a pronounced polarization dependence, enabling efficient electric-field control of the THz response.

In this Letter, we propose a new mechanism for controlling THz absorption and emission based on spin-corner-layer coupling (SCLC) in second-order topological altermagnetic bilayers [Fig. \ref{fig1}]. By using an electric field to influence electrons between different layers, this mechanism enables the simultaneous manipulation of both corner and spin d.o.f. Consequently, it allows for the control of the absorption, emission intensity, and even the polarization of THz waves. Through first-principles calculations, we identify a bilayer NiZrI$_6$ nanodisk---an intrinsic altermagnetic SOTI nanodisk---as a prototype platform exhibiting SCLC. The corners in NiZrI$_6$ nanodisks, which exhibit both opposing spin polarization ($P_s$) and layer polarization ($P_l$), facilitate the interaction among corner, layer, and spin d.o.f. A small out-of-plane electric field triggers dual switching of spin and layer polarizations of the corners, thereby enabling electric-field tuning of the THz responses. Our findings establish a viable route to electrically tunable THz spintronic platforms in 2D altermagnetic SOTI nanodisks.


\begin{figure}
\includegraphics[width=8.8cm]{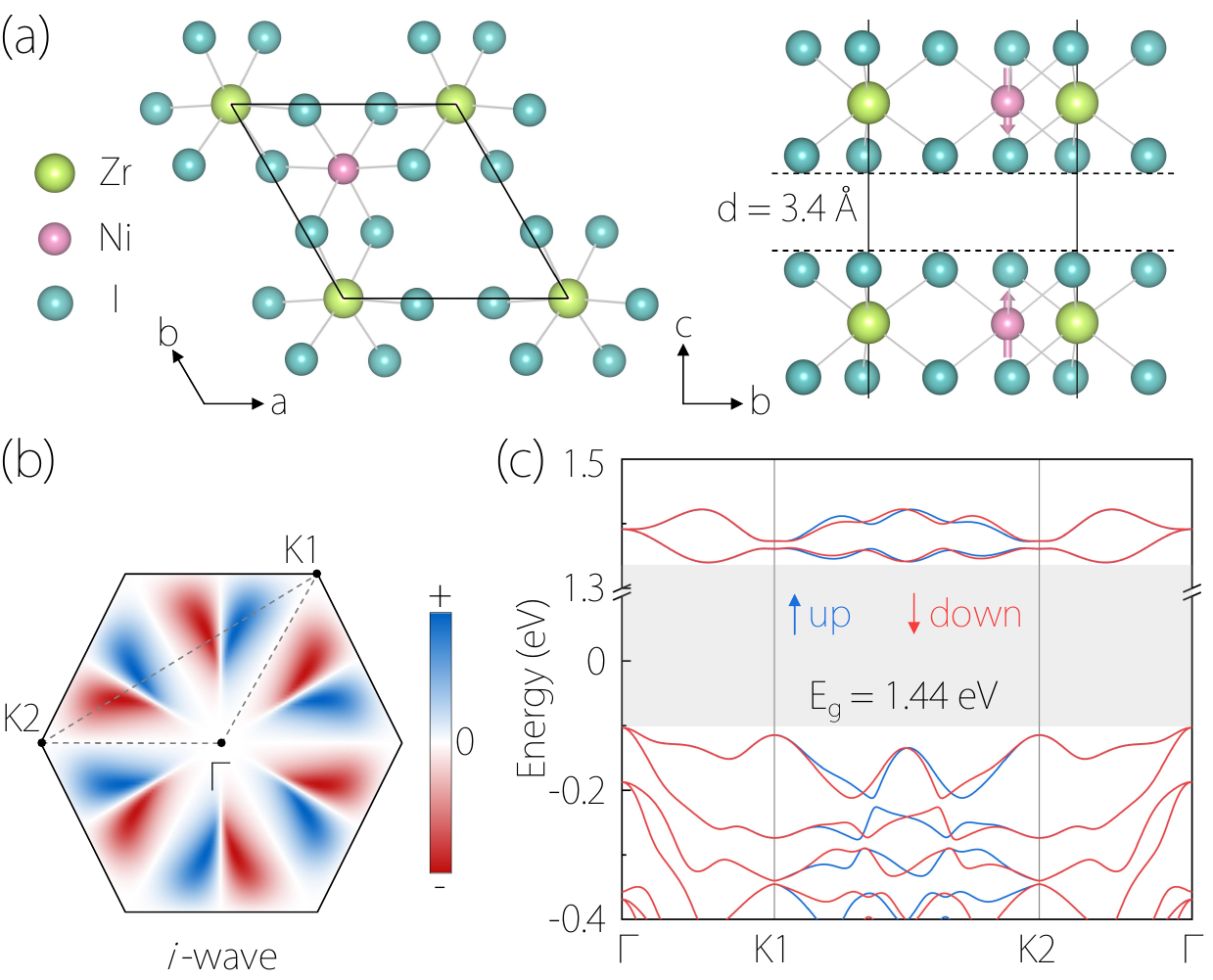}
\caption{(a) Crystal structure of bilayer NiZrI$_6$ with AA stacking, shown from different perspectives. (b) 2D surface Brillouin zone (BZ) and spin splitting of the valence band. (c) Spin-polarized band structure of the altermagnetic bilayer NiZrI$_6$ along the $\Gamma$-K1-K2-$\Gamma$ paths.
\label{fig2}}
\end{figure}

\begin{table}
\renewcommand\arraystretch{2}
\caption{\label{Tab1} The summary of rotation topological invariants and corner charge for altermagnetic bilayer NiZrI$_6$ with and without SOC effects. Here, $[\Pi_p^{(3)}]=\#\Pi_p^{(3)}-\#\Gamma_p^{(3)}~(p~=~1,~2,~3)$, which shows the difference between the number of states with $C_3$ eigenvalues at the $\Pi$ and $\Gamma$ points of the BZ.}
\begin{ruledtabular}
\begin{tabular}{cccccccc}
       & n$_{\rm{1a}}^{(\rm{ion})}$  & $\nu$ & [K$_1^{(3)}$] & [K$_1^{'(3)}$] & [K$_2^{(3)}$] & [K$_2^{'(3)}$] & $Q_{\rm{1a}}^{(3)}$ \tabularnewline
       \hline
      Spin- up & 12 & 64 & 1 & 1 & -1 & 0 & $\frac{|e|}{3}$ \tabularnewline
       \hline
       Spin-down & 12 & 64 & 1 & 1 & -1 & 0 & $\frac{|e|}{3}$ \tabularnewline
       \hline
       SOC & 24 & 128 & 0 & 1 & -1 & -1 & $\frac{2|e|}{3}$ \tabularnewline
\end{tabular}\end{ruledtabular}
\end{table}

\textcolor{blue}{\textit{SOTI bilayer with fractionally quantized corner charge}}---The bilayer NiZrI$_6$, belongs to space group $P312$ (No. 149), is predicted to exhibit dynamic stability and possesses an interlayer antiferromagnetic structure (see Fig.~\ref{fig2}(a))~\cite{48}. The optimized lattice constants are a = b = 7.19 \AA, and the Wyckoff positions of Ni, Zr, and I atoms are 2h, 2g, and 6l, respectively. The spin-polarized band structure for bilayer NiZrI$_6$ without spin-orbit coupling (SOC) is shown in Fig.~\ref{fig2}(c), in which the spin splitting appears on the K1-K2 path and a gap with the value of 1.44 eV occurs in both spin channels. Significantly, the in-plane twofold rotation symmetry $C_{2 \|}$ links two layers with opposing spins and is crucial for the emergence of $i$-wave altermagnetism in the bilayer NiZrI$_6$ (see Fig.~\ref{fig2}(b)). The computational methods can be found in the Supplemental Material (SM)~\cite{supp}.

Because the bilayer NiZrI$_6$ perseveres threefold rotation-symmetry $C_3$, and one can determine its topology by calculating the corner charge $Q_{\text {corner }}^{(3)}$ without considering the SOC, in terms of the rotation topological invariants, which shows as follow~\cite{52,53,54}.
{\small
\begin{equation}
Q_{1 a}^{(3)}=\frac{|e|}{3}\left(n_{1 a}^{(\mathrm{ion})}-\nu-\left[K_1^{(3)}\right]-\left[K_2^{(3)}\right]-\left[K_1^{\prime(3)}\right]-\left[K_2^{\prime(3)}\right]\right),
\end{equation}}
where $n_a^{(\text {ion })}$ is the ionic charges at Wyckoff position 1a, $\nu$ is the number of occupied bands. The summary of rotation topological invariants and corner charge $Q_{\text {corner }}^{(3)}$ are shown in Table~\ref{Tab1}. From it, one finds that the $Q_{\text {corner }}^{(3)}$ of a $C_3$-preserved NiZrI$_6$ nanodisk is $\frac{|e|}{3}$ for both spin-up and spin-down channels. The fractionally quantized corner charge reflects that the gaps in both spin channels are with nontrivial topology, and further, bilayer NiZrI$_6$ is an $i$-wave altermagnetic SOTI.

\begin{figure}
\includegraphics[width=8.8cm]{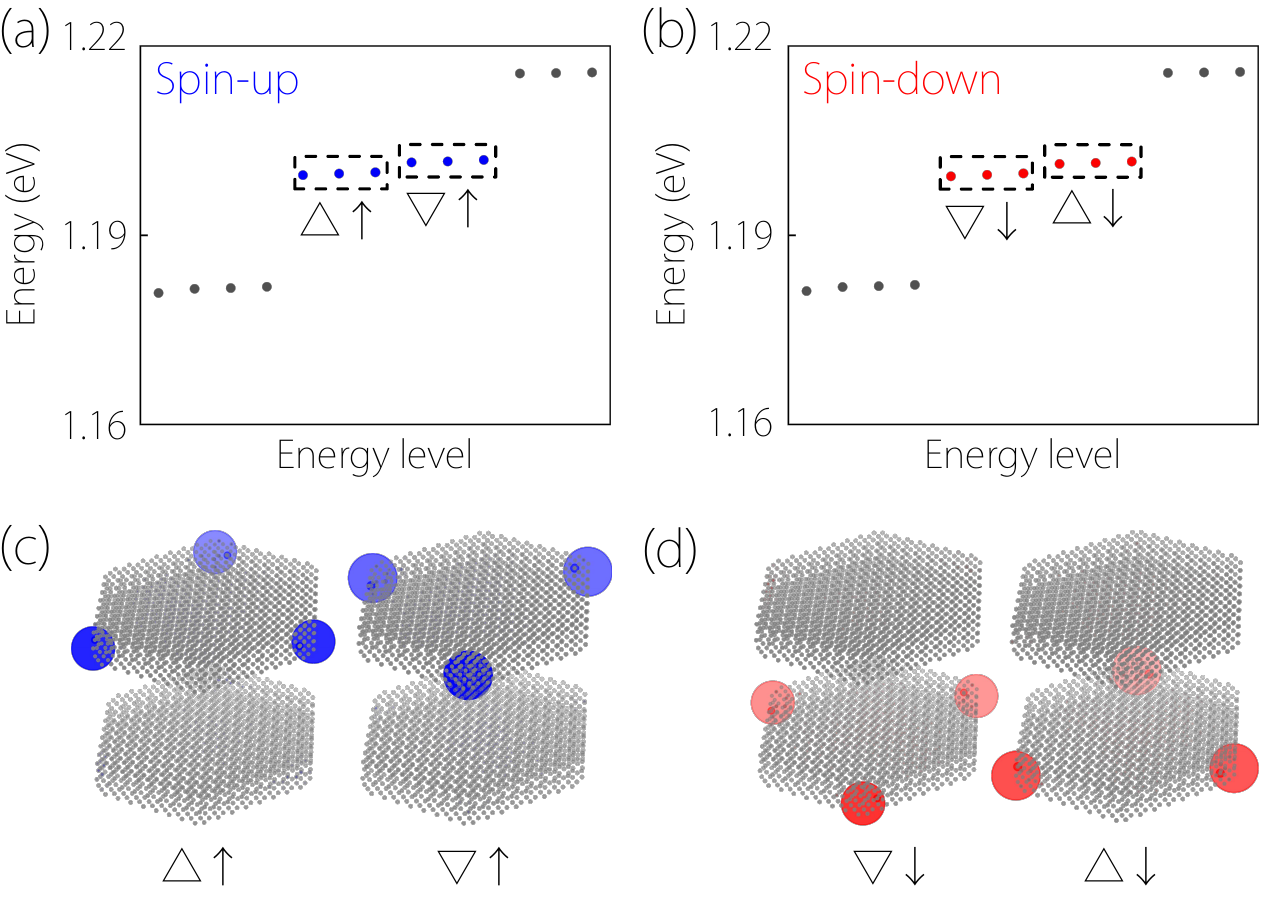}
\caption{(a)-(b) The energy spectrums for $C_3$-preserved NiZrI$_6$ nanodisk on a finite-size hexagonal nanodisk, where four groups of threefold degenerate states (the blue and red dots) can be observed. (c)-(d) The spatial distribution of the four groups of threefold degenerate states, termed as corners $\triangle\uparrow$, $\bigtriangledown\uparrow$, $\bigtriangledown\downarrow$, and  $\triangle\downarrow$, respectively. The $\triangle$ and $\bigtriangledown$ show that the corners are direction-dependent, and the $\uparrow$ and $\downarrow$ show that the corners have $P_l\left(P_s\right)>0$ and $P_l\left(P_s\right)<0$, respectively.
\label{fig3}}
\end{figure}

\begin{figure*}
\includegraphics[width=14cm]{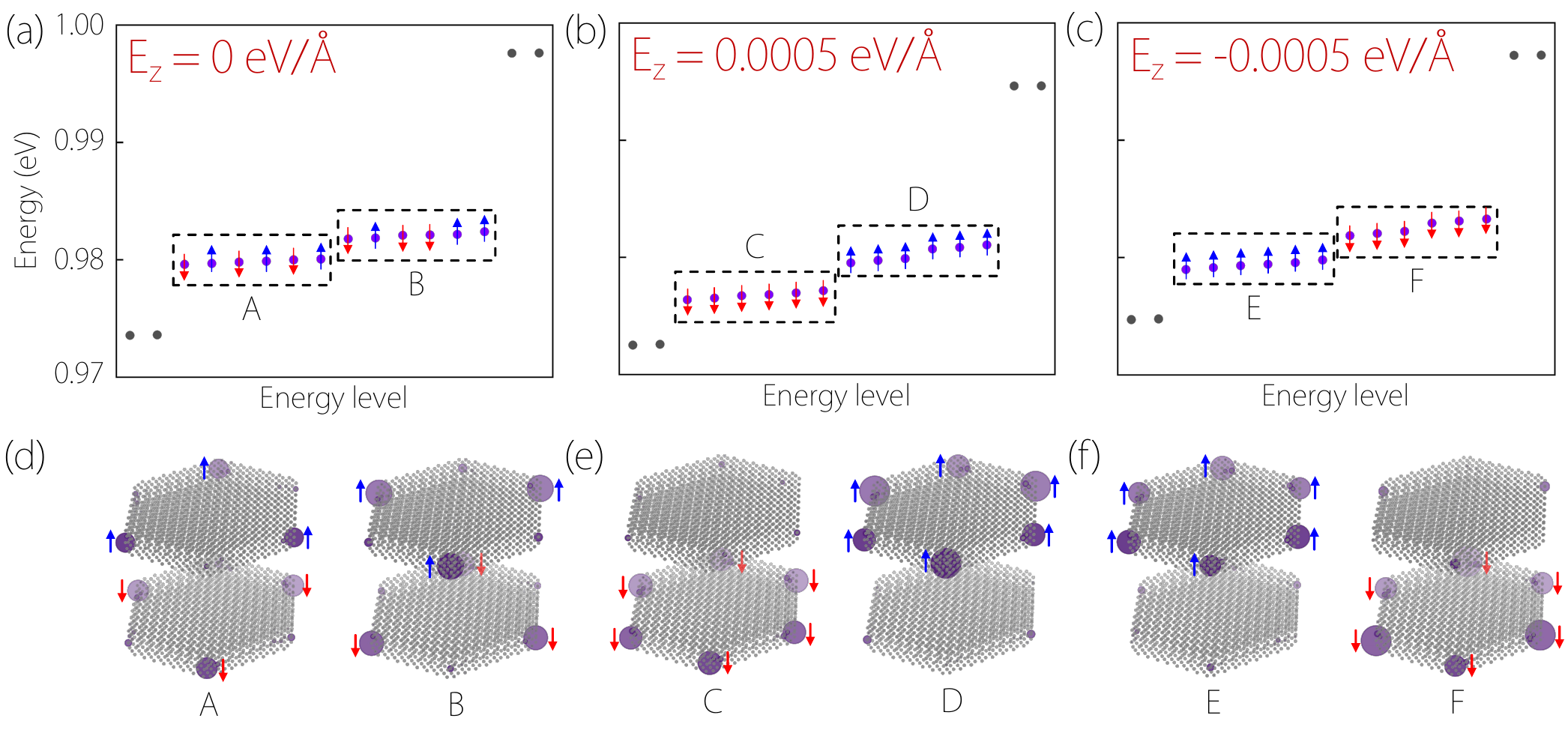}
\caption{(a)-(c) The energy spectrums for $C_3$-preserved NiZrI$_6$ nanodisk with SOC for $E_z$ = 0, $E_z$ = 0.0005, and $E_z$ = -0.0005 eV/\AA, respectively, in which six groups of sixfold degenerate states (A-F) can be observed. (d)-(f) The spatial distribution of the six groups of corner states, i.e., corners A-F.
\label{fig4}}
\end{figure*}

\textcolor{blue}{\textit{Spin-layer-coupled and direction-dependent corners}}---We come to study the corner states for the NiZrI$_6$ nanodisk in both spin channels. We construct a Wannier tight-binding model for a finite-size hexagonal nanodisk based on $C_3$-preserved bilayer NiZrI$_6$. The energy spectrums of the hexagonal nanodisk in both spin channels are shown in Fig.~\ref{fig3}. In Fig.~\ref{fig3}(a), two groups of threefold degenerate states (blue dots) appear within the bulk and surface states (black dots).
The spatial distributions for these two groups of  states are plotted in Fig.~\ref{fig3}(c). One can confirm that the threefold degenerate states are indeed corner states localized at three out of six corners of the finite-size hexagonal nanodisk. Interestingly, these two groups of corners are direction-dependent corners of the nanodisk, termed as corner $\triangle$ and corner $\bigtriangledown$, respectively.

Then, one can define the spin polarization and layer polarization of a corner state $|\psi(\boldsymbol{k})\rangle$ as
\begin{equation}
P_{s(l)}=\langle\psi(\boldsymbol{k})| \hat{P}_{s(l)}|\psi(\boldsymbol{k})\rangle ,
\end{equation}
with $\hat{P}_s \equiv \hat{s}_z\left(\hat{P}_l \equiv \hat{r}_z\right)$ denoting the $z$ component of the spin (position) operator. $P_l>0\left(P_s>0\right)$ shows that $|\psi(\boldsymbol{k})\rangle$ distributes more weight in the top layer (spin-up), while $P_l<0\left(P_s<0\right)$ shows that $|\psi(\boldsymbol{k})\rangle$ distributes more weight in the bottom layer (spin-down). Hence, corner $\triangle$ and corner $\bigtriangledown$ in Fig.~\ref{fig3}(c) are completely distributed in spin-up ($P_s>0$) and are situated in the top layer ($P_l>0$). They are therefore denoted as corner $\triangle\uparrow$ and corner $\bigtriangledown\uparrow$, respectively.

 As shown in Fig.~\ref{fig3}(b), the threefold degenerate states (red dots), i.e., the direction-dependent corners $\bigtriangledown$ and $\triangle$, are also present within the bulk and surface states (black dots).   {\color{black}The corner states in Fig.~\ref{fig3}(b)} are totally distributed in spin-down ($P_s<0$) and located in the bottom layer ($P_l<0$) (see Fig.~\ref{fig3}(d)), where they denote as corners $\bigtriangledown\downarrow$ and  $\triangle\downarrow$. In NiZrI$_6$ nanodisk, due to the altermagnetism, the direction-dependent corners in two layers with opposite $P_l$ must come in pairs with opposite $P_s$, which is indicative of the nature of spin-layer-coupled corner states and results in the SCLC effect.

Note that the hexagonal nanodisk constructed from a NiZrI$_6$ bilayer has a side length of $8a$ ($a = 0.72$~nm), corresponding to an area of approximately $86.20~\mathrm{nm}^2$. The SCLC effect remains robust against size reduction: when the side length is decreased from $8a$ to $4a$ (i.e., from $86.20~\mathrm{nm}^2$ to $21.55~\mathrm{nm}^2$), the spin-layer-coupled corner states still persist (see Figs.~S1 and S2 in the SM~\cite{supp}).

\textcolor{blue}{\textit{Electric-field-induced corners with dual-switchable $P_l$ and $P_s$}}---Notably, the SOTI nature and the spin-layer-coupled corners remain largely unchanged in NiZrI$_6$ nanodisk when SOC is incorporated. As shown in Table~\ref{Tab1}, under SOC effect, the $Q_{\text {corner }}^{(3)}$ = $\frac{2|e|}{3}$ reflects the robustness of the SOTI nature. Figure~\ref{fig4}(a) shows the energy spectrum NiZrI$_6$ nanodisk with SOC and without electric fields. One finds that two groups of sixfold degenerate states (A and B states) appear, and their spatial distributions are exhibited in Fig.~\ref{fig4}(d). Owing to the SOC, the $\triangle\uparrow$ and $\bigtriangledown\downarrow$ ($\triangle\downarrow$ and $\bigtriangledown\uparrow$) corners with opposite $P_l$$(P_s)$ coupled into corner A (B). By comparing with Figs.~\ref{fig3}(c), \ref{fig3}(d), and ~\ref{fig4}(d), the spatial distribution of $\triangle\uparrow$ and $\bigtriangledown\downarrow$ ($\triangle\downarrow$ and $\bigtriangledown\uparrow$) threefold degenerate states are almost the same with the spatial distribution of A (B) sixfold degenerate state, reflecting the robustness of the spin-layer-coupled corners in NiZrI$_6$ nanodisk under SOC.

Corner A (B) can be roughly viewed as the merger of corner $\triangle\uparrow$ ($\bigtriangledown\uparrow$) with $P_l$$(P_s)$$>0$ and corner $\bigtriangledown\downarrow$ ($\triangle\downarrow$) with $P_l$$(P_s)$$<0$, as illustrated in Fig.~\ref{fig4}(d). Interestingly, we can anticipate the realization of an electric-field-induced switchable sign of $P_l\left(P_s\right)$  for the corners as a result of the SCLC and direction-dependent corners. Under a positive electric field, such as $E_z$ = 0.0005 eV/\AA, the corner $\triangle\uparrow$ ($\triangle\downarrow$)  in corner A (B) changes to the corner $\triangle\downarrow$ ($\triangle\uparrow$), as illustrated in Figs.~\ref{fig4}(b) and~\ref{fig4}(e). Conversely, the corner $\bigtriangledown\downarrow$ ($\bigtriangledown\uparrow$) in corner A (B) has been inverted to the corner $\bigtriangledown\uparrow$ ($\bigtriangledown\downarrow$) under a negative electric field, such as $E_z$ = -0.0005 eV/\AA (see Figs.~\ref{fig4}(c) and~\ref{fig4}(f)).

Hence, one can conclude that the signs of $P_l$ and $P_s$ for the direction-dependent corners $\triangle$ and $\bigtriangledown$ can be dual-switched by reversing the electric field. The reasons can be explained as follows: Spin-layer locking enables the modulation of layers via electric fields, thereby controlling the spin orientation. The electric field breaks the $C_{2 \|}$ symmetry: a positive (upward) electric field raises the energy of the corner states in the upper layer, and a negative (downward) electric field elevates the energy of the corner states in the lower layer. As a result, the spin for the corners experiences a corresponding reversal in alignment with the reversal of the layer for the corners.

\begin{figure}
\includegraphics[width=8.8cm]{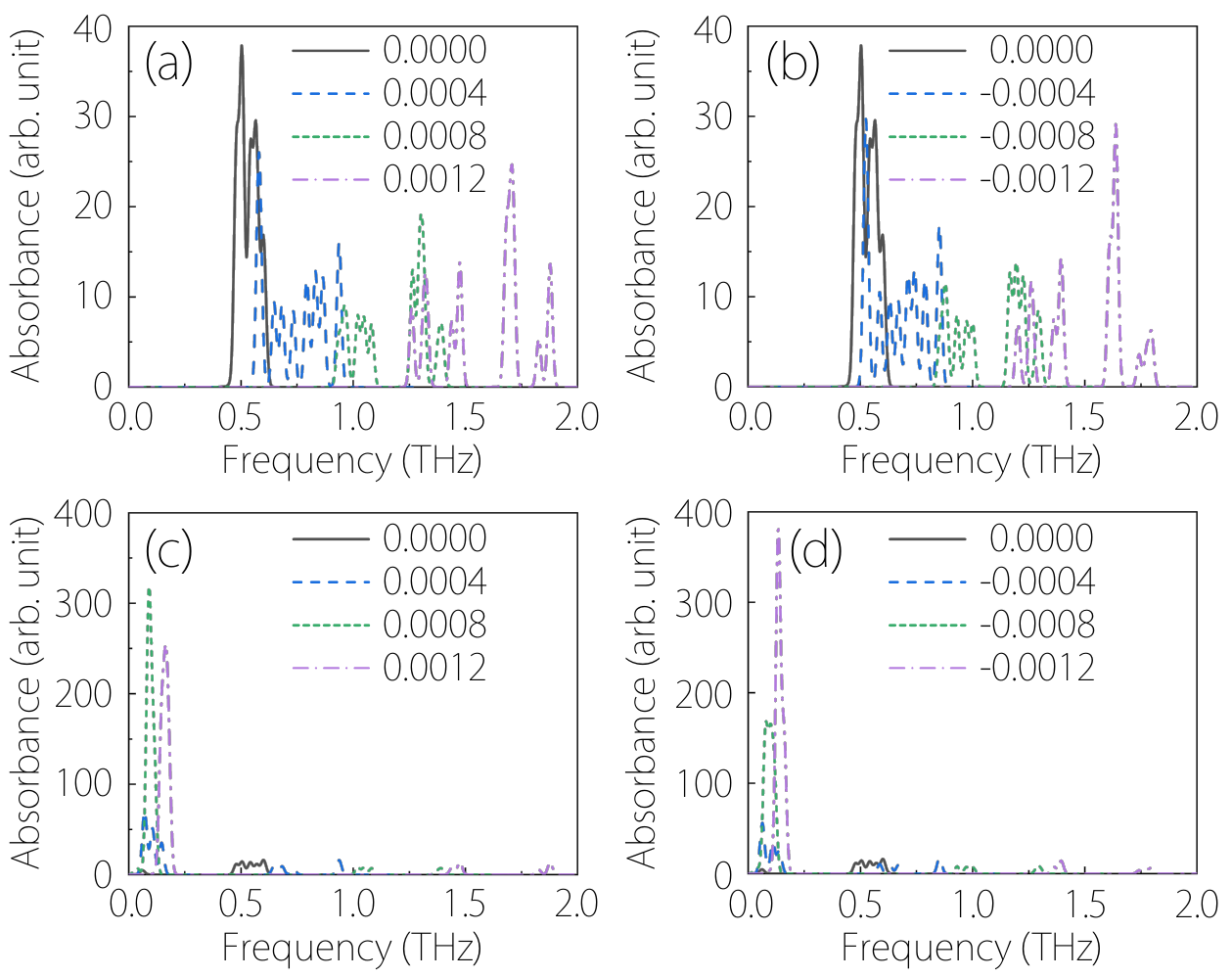}
\caption{(a)-(b) Absorption spectrum when the highest occupied state is at the sixth corner state. (c)-(d) Absorption spectrum when the highest occupied state is at the third corner state. Black solid lines denote the case without an electric field, whereas colored dashed lines represent  the cases with different values of positive (a, c) and negative (b, d) electric fields.
\label{fig5}}
\end{figure}

\textcolor{blue}{\textit{Electric-field tuning of the THz response}}---The corner‐state energy gap of NiZrI$_6$ nanodisk, a type of QD,  is only a few meV, indicating that it can correspond to electromagnetic waves in the THz frequency range.
{\color{black}Such behavior is unique to corner-state QDs, whereas conventional nanometer-sized QDs usually exhibit gaps much larger than the THz energy scale~\cite{55}.}

In spectroscopy, {\color{black}the THz responses can be quantitatively described by the absorbance.}
The absorbance $f_{12}$ of a transition from a state $|\psi_1\rangle$ to a state $|\psi_2\rangle$ is defined as
\begin{equation}
       f_{12} = \frac{2}{3} \frac{m_e}{\hbar^2} \left(E_2 - E_1\right) \sum_{\alpha = x, y, z} \left| \left\langle \psi_1 \middle| R_\alpha \middle| \psi_2 \right\rangle \right|^2,
\end{equation}
where $m_e$ is the mass of an electron and $\hbar$ is the reduced Planck constant. $E_1$ and $E_2$ are the energies of states $|\psi_1\rangle$ and $|\psi_2\rangle$, respectively. The $\left\langle \psi_1 \middle| R_\alpha \middle| \psi_2 \right\rangle$ is the transition dipole moment of two states, where $R_\alpha$ is the position operator of $\alpha$ direction. Here, we only consider the absorbance between the corner states since the transitions between other states are out of the range of THz frequency. To simulate the experimentally observed absorption spectrum, we apply broadening to the theoretically obtained transition data, resulting in absorbances manifested as a series of absorption peaks.

We first consider the scenario where electrons occupy half (one-sixth) of the twelve available corner states. The absorption peak is observed near 0.5 THz, as shown in Figs. \ref{fig5}(a) and \ref{fig5}(b). In this configuration, both the occupied and unoccupied corner states are distributed across the upper and lower layers [Fig. \ref{fig4}(d)]. This allows for intralayer transitions, resulting in a large oscillator strength and a prominent absorption peak.
However, when an upward or downward electric field is applied, the occupied corner states become exclusively distributed in either the upper or lower layer, while the unoccupied states are confined to the opposite layer [Figs. \ref{fig4}(e), \ref{fig4}(f), and S3~\cite{supp}]. This spatial separation suppresses intralayer transitions, leading to a significant reduction in oscillator strength and a substantial decrease in the absorption peak's intensity. Concurrently, the increasing energy gap between the occupied and unoccupied corner states causes a blueshift of the absorption peak. At an electric field strength of $\pm$ 0.0012 eV/\AA, the peak shifts to nearly 2 THz, as depicted in Figs. \ref{fig5}(a) and \ref{fig5}(b).

Next, we examine the scenario where electrons occupy one-third of the corner states. The absorption peak remains at 0.5 THz, as shown in Figs. \ref{fig5}(c) and \ref{fig5}(d). In contrast to the half-filling case, the application of an electric field produces the opposite effect. The absorption peak intensity increases significantly, while its position redshifts to approximately 0.1 THz [Figs. \ref{fig5}(c) and \ref{fig5}(d)]. This occurs because the occupied and unoccupied corner states remain distributed within the same layer [Figs. \ref{fig4}(e), \ref{fig4}(f), and S3~\cite{supp}]. For instance, with an upward electric field (positive value), the three occupied corner states are all in the lower layer, along with three of the unoccupied states. This co-location enables intralayer transitions, leading to a notable enhancement of the absorption peak. Simultaneously, the reduced energy gap between the occupied and unoccupied states causes the observed redshift.
Furthermore, with an upward electric field, the absorption peak is primarily contributed by spin-down electrons, whereas a downward electric field leads to a contribution from spin-up electrons [Fig. S3 in the SM~\cite{supp}]. Therefore, while the position and intensity of the absorption peak remain relatively unchanged when the electric field direction is reversed, the different spins contributing to the absorption result in a change in the direction of the transition dipole moment [Table S1 in the SM~\cite{supp}]. This allows the material to exhibit a different absorption capability for different polarizations of THz electromagnetic waves when the electric field is switched.

\textcolor{blue}{\textit{Summary}}---In this Letter, we propose a new concept of coupling among corner, spin, and layer d.o.f., named SCLC. The SCLC originates from the spin-layer-coupled corner states, distinguished by corners exhibiting opposing $P_l$ and $P_s$. Based on a first-principle calculation, we predict that the nanodisk constructed from altermagnetic bilayer NiZrI$_6$ with AA stacking is the first platform with SCLC. Due to the SCLC in the NiZrI$_6$ nanodisk, reversing the direction of the external electric field can dual-switch the signs of $P_l$ and $P_s$ of the corners, thereby enabling effective control of THz response. The SCLC effect remains robust under size reduction, and the NiZrI$_6$ nanodisk operates entirely at the nanometer scale, offering a compact and electrically driven platform for THz spintronics. This substantial contribution has the potential to significantly advance the field of 2D SOTIs, altermagents, corner-state QDs, and multi-d.o.f. couplings, as well as the development and application of nanoscale THz spintronic platforms.
\bibliography{reference}

\end{document}